\documentclass{aa}
\input epsf 
\usepackage{times}
\begin{document}

   \thesaurus{06     
              (08.02.1; 08.09.2 Aql X-1; 08.14.11; 13.25.5)}
   \title{Magnitude, color and spectral type of 
   Aql X-1 in quiescence
   \thanks{Based on observations obtained at the European Southern Observatory,
   La Silla, Chile}}

   \author{Claude Chevalier\inst{1}, Sergio A. Ilovaisky\inst{1},
          Pierre Leisy\inst{2}  
         \and Ferdinando Patat\inst{2}}

   \offprints{C.Chevalier}

   \institute{Observatoire de Haute-Provence (CNRS), F-04870 St.Michel l'Observatoire, France\\
              email: chevalier, ilovaisky@obs-hp.fr
                    \and
            European Southern Observatory, Casilla 19001, Santiago 9, Chile\\
             }

   \date{Received 1999 May 14; accepted 1999 June 16}
   \authorrunning{C.Chevalier, S.A.Ilovaisky, P.Leisy and F.Patat}
   \titlerunning{Aql X-1 in quiescence}
   \maketitle

   \begin{abstract}

   Direct $I$ and $V$-band imaging of Aql X-1 in quiescence and during outburst maximum shows that the true optical counterpart is the interloper located 0.48\arcsec\ West of the previously known star. We find for the new counterpart $V$ = 21.6 and $V-I$ = 2.2 in quiescence, when its contribution to the total light was 12\% in $V$ and 22\% in $I$.  Analysis of this photometry and of low-resolution spectra of the sum of both stars, also taken during quiescence, shows that the likely spectral types for the previously known star and the optical counterpart are late G and late K, respectively, reddened by $E(B-V)$ = 0.5 $\pm$0.1. 

      \keywords{binaries: close --
                stars: individual: Aql X-1 --
                stars: neutron --
                X-rays: stars 
               }
   \end{abstract}

%

\section{Introduction}

   Recent $K$-band imaging with Keck I of the transient low-mass X-ray binary Aquila X-1 (= V1333 Aql) has resolved V1333 Aql into two stars lying approximately along the east-west direction and separated by 0.46\arcsec\ (Callanan et al. \cite{Call}), the easterly star contributing 60\% of the combined flux at $K$. Observations in the $z$-band (1.05$\mu$) indicated also that the easterly star was "somewhat bluer" and they speculated that it might be the true optical counterpart of Aql X-1.

In 1999 we obtained $V$ and $I$-band frames of Aql X-1 both in quiescence and during outburst maximum with EFOSC-2 at the ESO 3.6-m telescope. We present here the results of our photometry and of our EFOSC-1 spectroscopy, and we  show that the westerly star is the true counterpart.


   	\begin{figure*}
   	\epsfxsize=18cm
	\epsfbox{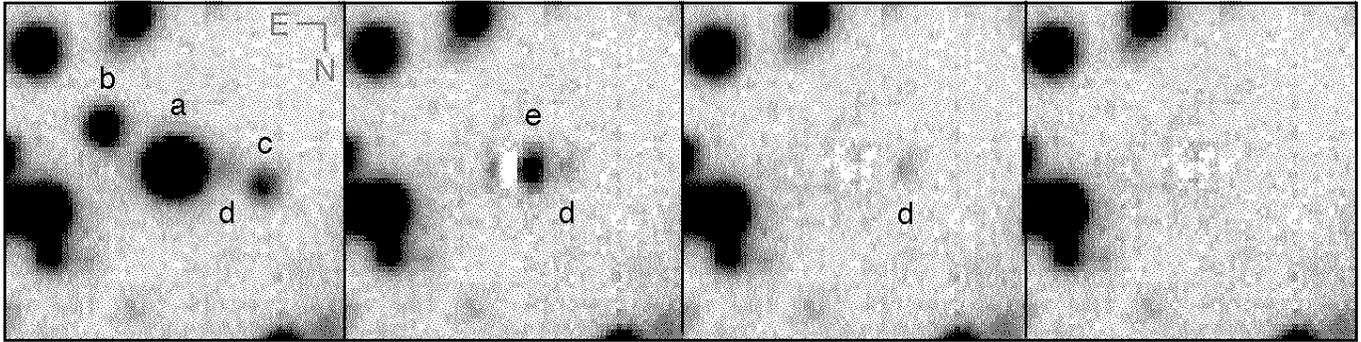}
    \caption{
{\bf (a) Left panel:} A close-up view of our medianed $I$-band frame, taken with EFOSC-2 at the ESO 3.6-m telescope on 1999 April  9, showing a 10\arcsec$\times$10\arcsec\ field around Aql X-1 (0.157\arcsec/pixel). The FWHM for the image profiles is 0.8\arcsec. Star $a$ was previously believed to be the optical counterpart of this recurrent transient. The image is reversed with North at the bottom and East to the left.
{\bf (b) Second panel:} The resulting image after subtraction of the cleaned third-generation PSF profile fitted to stars $a$, $b$ and $c$. A new object, star $e$, is now clearly visible but the residuals indicate a poor over-all fit due to the presence of star $e$.
{\bf (c) Third panel:} Here star $e$ has been included in the fitting process and subtracted out with an immediate improvement in the residuals. Only star $d$ remains.
{\bf (d) Right panel:} All stars have now been removed and the residuals are very small.}
	\end{figure*}


\section{Photometry}
\subsection{Quiescence}
Three dithered 5-min $V$-band exposures and 3 dithered 5-min $I$-band exposures (using actually a Gunn $i$ filter) were obtained on 1999 April 9 with the imaging spectrograph EFOSC-2 at the Cassegrain focus of the ESO 3.6-m telescope at La Silla, using a Loral/Lesser CCD (\#40) with 15$\mu$ pixels, yielding a projected pixel size of 0.157\arcsec. Conditions were clear, the seeing was 1\arcsec\ for the $V$ frames and 0.8\arcsec\ for the $I$ frames. After standard MIDAS reduction procedures, the medianed $V$ and $I$ frames were measured using the DAOPHOT {\it NSTAR} routine of Stetson (\cite{Stet}). A clean point spread function (PSF) profile was derived from three nearby isolated stars after three neighbour-removing iterations.  Two comparison stars, C1 and C2, situated (12.7\arcsec N, 8.7\arcsec S) and (14\arcsec N, 25\arcsec S) respectively from Aql X-1, were used to derive the $V$ and $I$ magnitudes. Their Johnson-Cousins magnitudes, measured at Observatoire de Haute-Provence with the 1.2-m telescope,  are $V$ = 17.48, $I$ = 16.12 for C1 and $V$ = 17.42, $I$ = 16.29 for C2 (Chevalier and Ilovaisky \cite{CIc}). We estimate the accuracy of these magnitudes to be 0.05 mag in each band.

\begin{table*}
\renewcommand{\footnoterule}{\rule{5mm}{0mm}\vspace{-5mm}}
\caption[]{Photometric results from PSF fitting using DAOPHOT $NSTAR$\\}
\begin{flushleft} 
\begin{minipage}{10cm}
\vspace{-\abovedisplayskip}
\begin{tabular}[b]{ccccccc}
	\hline
Star & separation & 
$V$ & $I$ & $V-I$ & $(V-I)_{\circ}$\footnote{de-reddened using $E(V-I) = 0.65$ (see Discussion)} & Sp.Type\footnote{
derived from the colors computed by Bessell (\cite{Bess}) for Vilnius spectra}  \\
 & (\arcsec) & (mag) & (mag) & (mag)	& (mag) & (Class V) \\
	
	\hline
a & 0.00 & 19.42 $\pm$ 0.06 & 18.02 $\pm$ 0.06 &
1.40 $\pm$ 0.09 & 0.75 & F8-K0\\

b &
2.30&
21.17 $\pm$ 0.06&
19.66 $\pm$ 0.06&
1.51 $\pm$ 0.09&
0.86 &
G6-K2\\

c&
2.87&
22.75 $\pm$ 0.07&
20.62 $\pm$ 0.06&
2.13 $\pm$ 0.10&
1.48 &
K5-K7\\

d&
1.78&
24.63 $\pm$ 0.34&
22.07 $\pm$ 0.09&
2.56 $\pm$ 0.35&
1.91 &
K7-M3\\

e&
0.48&
21.60 $\pm$ 0.10&
19.37 $\pm$ 0.07&
2.23 $\pm$ 0.12&
1.58& 
K6-M0\\

	\hline
\end{tabular}
\end{minipage}
\end{flushleft}
\end{table*}

A close-up view (10\arcsec$\times$10\arcsec) in the $I$-band of the stars around Aql X-1 (star $a$) is shown in Fig. 1a while Fig. 1b shows the same image after subtraction of the PSF profile at the location of stars $a$, $b$ and $c$, revealing the interloper star $e$, in addition to star $d$. The {\it NSTAR} fitting process, which treats the positions as input parameters to be optimized, was incomplete at this point as the subtracted image shows strong residuals, including star $e$. An estimate of the goodness of fit can be derived from the parameters CHI and SHARP produced by the {\it NSTAR} routine. CHI is the ratio of the observed pixel-to-pixel scatter in the fitting residuals to the expected scatter based on the values of read-out noise and gain, and should not exceed unity if the fit is good. SHARP  measures the difference between the half-width at half-maximum for a star and that for the PSF and is close to zero for isolated stars.  When only stars $a$, $b$ and $c$ were included in the fit (Fig 1b), star $a$ yielded CHI = 4.62 and SHARP = 0.027, indicating a poor fit.

Fig. 1c shows star $d$ which is left after subtraction of stars $a$, $e$, $b$, $c$ and Fig. 1d shows the cleaned image after removing the profiles fitted to $a$, $b$, $c$, $d$, and $e$.  Inclusion of star $e$ (Fig 1d) in the fit yielded CHI = 0.85 and SHARP = 0.007 for both stars $a$ and $e$, showing a good fit. The same procedure was applied to the medianed $V$ frame. The final results of the {\it NSTAR} photometry are given in Table 1. The main sources of the errors given in Table 1 are the uncertainty on the absolute calibration (zero-point) of the comparison stars C1 and C2 (color terms are relatively small) and the error on the {\it NSTAR} fitting for stars $e$ and $d$. These determinations are of much better quality than our previous estimates (Chevalier and Ilovaisky \cite{CIa}) based on the 1989 $V$-band frames which were obtained with a seeing of 1.2-1.3\arcsec\ and a projected pixel size of 0.33\arcsec. On these frames star $d$ was undetected and the magnitude of star $e$, which appeared as a faint residual after {\it NSTAR} fitting to the $a$, $b$, $c$ group of stars, was underestimated. Under these conditions, the measured magnitude for star $a$, $V$ = 19.26, is an overestimate by 0.1-0.15 magnitude due to the contamination by star $e$ and by the surrounding objects. Measurements obtained with smaller telescopes are affected in a similar fashion, depending on the projected pixel size and seeing.

   	\begin{figure}
   	\epsfxsize=4.5cm
	\epsfbox{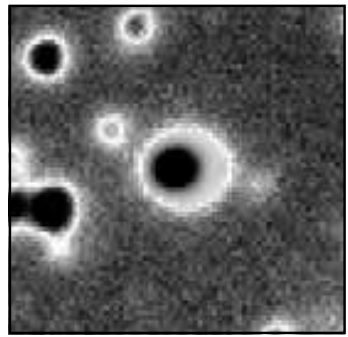}
    \caption{
The 1-min $I$-band CCD image obtained near outburst maximum on 1999 May 21  with EFOSC-2 is shown as a positive print and the image corresponding to quiescence (Fig. 1a) has been aligned and overlayed as a negative print (field size and orientation are the same as in Fig 1a). The seeing for the two frames was different. Note how the image of the object in outburst is clearly shifted to the West relative to the image in quiescence, showing that star $e$ of Fig. 1b is the true optical counterpart of Aql X-1.
}
    \end{figure}


\subsection{Outburst}
During early 1999 May, the source started a new outburst (Jain et al. \cite{Jain}, Chevalier and Ilovaisky \cite{CIb}) and on May 21, near outburst maximum, we secured  1-min CCD frames in $V$ and $I$ with EFOSC-2. The FWHM of the image profiles was 1.1\arcsec\ in $I$. Inspection of the frames  showed that the barycenter of the variable object image in outburst did not coincide with the position of star $a$.  We analyzed  the outburst frames with DAOPHOT using as a starting point the table of star positions derived by {\it NSTAR} from the quiescent frames. From the positions of eight near-by objects we
find that the variable object is located  $-$0.05 $\pm$0.03\arcsec\ in right-ascension and $-$0.03 $\pm$0.06\arcsec\ in declination from the position of star $e$, which is located 0.48\arcsec\ West of star $a$ (Table 1). The lower signal-to-noise ratio in these short exposures did not
allow accurate photometry of star $a$ but subtraction of a PSF profile fitted to star $e$ shows a residual compatible with star $a$. On these images, star $e$ has $V$ = 17.03 and $V-I$ = 1.03, with a slight contamination from $a$. Fig. 2 shows the same field of view as  Fig. 1 with the outburst frame shown as a positive print and the quiescent frame overlaid as a negative print. The  image in activity appears shifted by 0.42 $\pm$0.03\arcsec\ to the West from the position of star $a$, as measured in quiescence.


   	\begin{figure}

   		\epsfxsize=8.7cm
	\epsfbox{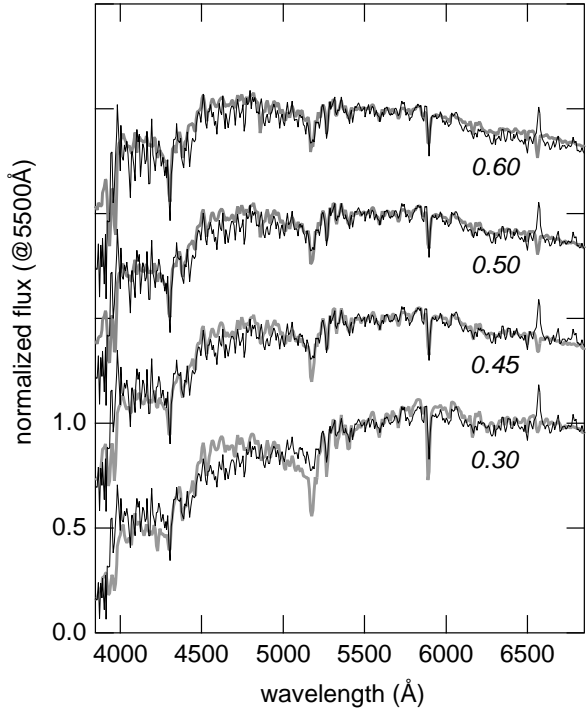}
    \caption{
Results from ESO 3.6-m EFOSC-1 spectra taken in 1988 May 19 and 1989 May 8.  The spectrum of the sum of stars $a$ and $e$ is plotted corrected for different values of the reddening, $E(B-V)$ = 0.30, 0.45, 0.50 and 0.60, from bottom to top, and is normalized to 1.0 at 5500\AA. The top three curves are each shifted vertically by 0.5 from the previous one for clarity. 
Also plotted are the different combinations, at 5500\AA\ as described in the text, of 88\% of the spectrum of an earlier type star (K3, K0, G8 and G5, bottom to top) and 12\% of the spectrum of a K7V star (G~747.3), all taken with the same instrument. 
}

    \end{figure}


   	\begin{figure}
   	\epsfxsize=8.7cm
	\epsfbox{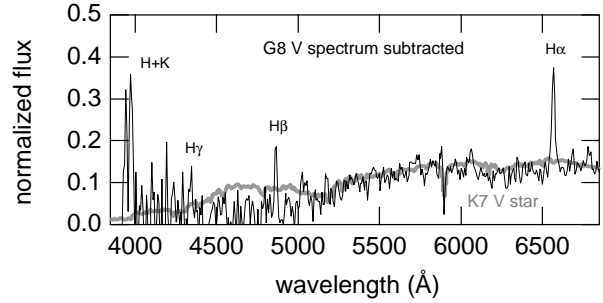}
    \caption{
Result of the subtraction of the synthetic G8V spectrum  (derived from our observed spectra as described in the text) from the sum of stars $a$ and $e$, dereddened by $E(B-V) = 0.50$. Also plotted is the spectrum of a K7 V star (grey line), normalized to 12\% of the sum at 5500\AA\ (See Fig. 3). The main  emission lines are identified.
}
   \end{figure}

\section{Spectroscopy}

 We obtained several spectra of Aql X-1 in quiescence with the imaging-spectrograph EFOSC-1 at the Cassegrain focus of the ESO 3.6-m telescope. They were taken through the B300 grism and cover the range between 3850\AA\ and 6850\AA. Two spectra were obtained during the night of 19 May 1988 with 7\AA\ resolution (using CCD \#11) and exposure times of 3600s and 4200s. Two more spectra were obtained one year later on 8 May 1989 with 3.5\AA\ resolution (using CCD \#8) and exposure times of 3600s and 2700s. The second 1989 spectrum was underexposed and will not be discussed further. Weather conditions were good throughout, with seeing of 1.1\arcsec\ and 1.3\arcsec\ respectively, and we used a slit width of 1.5\arcsec. Spectra of bright stars of known spectral type from mid-GV to early MV (G3: HD~168402, G5: G~724.1, K0: HD~171982, K3: HD~87521, K7: G~747.3, M0: CD~$-$36\degr6589) were also taken during the 1988 run.
  All frames were wavelength-calibrated using He-Ar lamp spectra and flat-fielded using internal Tungsten lamp spectra and were reduced using standard MIDAS procedures. The spectra were extracted with the MIDAS long-slit package and corrected for atmospheric extinction using average values for La Silla. The standard star LTT 7987 (Stone and Baldwin \cite{Sto}) was used for relative flux calibration. The three spectra of Aql X-1, which did not show any significant differences, were then co-added.  Since our observed G8 V star (CD~$-$45\degr12143) turned out to be of a K0 type, we constructed a synthetic G8 V spectrum by averaging our G5 and K0 spectra.

The slit included both stars $a$ and $e$. As determined using Table 1, the flux coming from star $e$ is 12\% of the total flux from stars $a$ and $e$ in the $V$-band. To derive the spectral type of star $a$, we constructed a set of template spectra by adding, to each of our observed spectra of known spectral type, normalized to 0.88 at 5500\AA, the observed K7 or M0 spectrum, normalized to 0.12 at 5500\AA\ (see Table 1 and also the Discussion). We applied different corrections for interstellar reddening to the ($a$+$e$) spectrum, corresponding to $E(B-V)$ from 0.3 to 0.6  and normalized the de-reddened ($a$+$e$) spectra to unity at 5500\AA. This assumes that both stars are reddened by the same amount (see Discussion).
We then compared the set of de-reddened observed ($a$+$e$) spectra to the set of template spectra. The template spectra including an M0 component were rejected since they exhibited residual TiO bands which do not appear in the observed ($a$+$e$) spectrum and thus are not shown here. The absorption features of the observed ($a$+$e$) spectrum, in particular the G-band of CH ($\lambda\lambda$4290-4314\AA), are only compatible with a late G or early K type for star $a$.  

Fig. 3 shows four different template  spectra (K3+K7, K0+K7, G8+K7 and G5+K7) (grey curves) together with the average ($a$+$e$) spectrum de-reddened with $E(B-V)$ = 0.3, 0.45, 0.50 and 0.60, values selected to give a best fit to the template spectra. The depth of the Mg$b$ band at $\lambda$5175\AA\ (which was the only late-type feature in the first quiescent spectrum published by Thorstensen et al. \cite{Thor}) and of the TiO band at $\lambda$4954\AA\  increase with later spectral types and are too strong in the K3+K7 combination.  The ($a$+$e$) de-reddened energy distributions  for $E(B-V)$ = 0.60, 0.50 and 0.45 match almost equally well the G5+K7, G8+K7 or the K0+K7 composite templates, respectively, although there are differences at the blue and red ends.

In Fig. 4 we show the result of subtracting the synthetic G8 spectrum, assigned to star $a$, from the ($a$+$e$) spectrum, de-reddened for $E(B-V)$ = 0.50. This should approximate the spectrum of star $e$. Also shown (grey line) is our observed K7 V spectrum, normalized to 0.12 at 5500 \AA.
The main features of this difference spectrum are the emission lines at H$\alpha$, H$\beta$ (weaker) and Ca II K and H(+H$\epsilon$), which appear strong although the signal-to-noise ratio below 4500\AA\ in our spectra is low. Emission at H$\gamma$ is barely detectable and no emission is detected at \ion{He}{ii} $\lambda$4686\AA.  The slight deficit between 4400 and 5000 \AA\ may be of instrumental origin.

Our spectrum is different from that taken by Garcia et al. (\cite{Gar}) when the sum of Aql X-1 plus star $a$ was more than half a magnitude above quiescence ($V$ = 18.68), and which displays a richer emission-line spectrum, including \ion{He}{ii} $\lambda$4686\AA, a consequence of X-ray heating.

\section{Discussion}
 
 The absorption features present in our spectrum of the sum ($a$+$e$) indicate a spectral type for star $a$ between mid-G and K0, reddened by $E(B-V)$ = 0.5 $\pm$0.1. Using $E(V-I) = 1.3 \times E(B-V)$ (Dean, Warren and Cousins \cite{Dean}) and the same amount of reddening $E(V-I) = 0.65 \pm0.1$ for stars $a$, $b$, $c$, $d$ and $e$, we obtain the estimates, listed in the two right-hand columns of Table 1, for the de-reddened color index $(V-I)$ and the corresponding ranges of spectral types using the colors computed by Bessell (\cite{Bess}) for Vilnius spectra.

The magnitudes listed in Table 1 correspond to the flux ratios $f(e)/f(a) = 0.135$ or $f(e)/f(e+a) = 0.12$ in the $V$-band and $f(e)/f(a) = 0.29$ or $f(e)/f(e+a) = 0.22$ in the $I$-band. According to Callanan et al. (\cite{Call}), star $a$ contributes 60\% of the combined flux from the pair ($a$+$e$) at $K$, which gives a flux ratio $f(e)/f(a) = 0.67$ and a magnitude difference ${\Delta K}_{e-a} = K(e)-K(a) = 0.45$. In the $V$-band we find ${\Delta V}_{e-a} = V(e)-V(a) = 2.18$. Combining both results yields the difference in $V-K$ color between the two stars, ${\Delta (V-K)}_{e-a} = 1.73$. 

Assuming  a similar amount of reddening for stars $e$ and $a$, we derive the intrinsic $(V-K)$ color and spectral type for star $e$ as a function of the intrinsic $(V-K)$ colour and spectral type assumed for star $a$, using the main sequence calibration of Johnson (\cite{John}), and these are shown in Table 2. These determinations are compatible with the $V$ and $I$ photometric results of Table 1 and with a color excess $E(B-V)$ = 0.5 $\pm$0.1.  Assuming absolute visual magnitudes of ${M}_V = +5.8$ and $+8.1$ for stars $a$ and $e$ (Gray \cite{Gray}) and taking ${A}_V = 3.1 \times E(B-V)$, the magnitudes of Table 1 yield distances of 2.6 and 2.5 kpc, respectively, in agreement with the assumption of equal amounts of reddening for both objects. For a galactic latitude of $-4.1\degr$, such distances put the objects 180 pc below the galactic plane, beyond most of the absorbing layer. The low-energy X-ray absorbing column densities in the line of sight to Aql X-1 reported by Verbunt et al. (\cite{Verb}) and Zhang et al. (\cite{Zhang}) correspond to color excesses $E(B-V)$ of 0.59 $\pm$0.1 and 0.49 $\pm$0.1, respectively (using the relation of Ryter et al. \cite{Ryter}). These values are compatible with those  used here.
\begin{table}
\caption[]{Derived colors and spectral types for star $e$ from assumed values for star $a$}
\begin{flushleft} 
\begin{minipage}{8.7cm}
\begin{tabular}[b]{cccc}
	\hline
\multicolumn{2}{c} {star $a$}  &  
\multicolumn{2}{c} {star $e$}  \\  
	\hline
assumed  & assumed & derived & derived\\ 
Sp.Type & ${(V-K)}_{\circ}$ & ${(V-K)}_{\circ}$ & Sp.Type\\
	\hline
G5 & 1.49 & 3.22 & K7 \\
G8 & 1.63 & 3.36 & K7-K8 \\
K0 & 1.83 & 3.56 & K8-M0 \\
	\hline
\end{tabular}
\end{minipage}
\end{flushleft}
\end{table}

\section{Conclusions}
We have shown that the true optical counterpart of the Aquila X-1 recurrent transient is the interloper reported by Callanan et al. (\cite{Call}), located 0.48\arcsec\ West of the star previously assumed to be the candidate, now an ordinary mid to late-G type star. Photometry and low-resolution spectroscopy obtained in quiescence give for the new counterpart $V$ = 21.6 and $V-I$ = +2.2 and suggest a star of K7 V spectral type, reddened by $E(B-V)$ = 0.5 $\pm$0.1 and located at 2.5 kpc. This makes Aql X-1 similar to most other soft X-ray transients which have late K (or M0) companions. We defer  examination of the consequences of this new identification on existing optical photometry to a forthcoming article (Chevalier and Ilovaisky\cite{CIc}).


\end{document}